# INERTIAL MIGRATION OF NEUTRALLY BUOYANT PARTICLES IN SQUARE CHANNELS AT HIGH REYNOLDS NUMBERS

**Yichang Wang[1], Yanfeng Gao[1], Pascale Magaud[1,2], Lucien Baldas\*[1], Christine Lafforgue[3], Stéphane Colin[1]**

[1]Institut Clément Ader (ICA), Université de Toulouse, INSA, ISAE-SUPAERO, Mines-Albi, UPS,
3 rue Caroline Aigle, 31400 Toulouse, France
yichang.wang@insa-toulouse.fr, yanfeng.gao@insa-toulouse.fr, lucien.baldas@insa-toulouse.fr,
stephane.colin@insa-toulouse.fr

[2]Université de Limoges, 33 rue François Mitterrand, 87032 Limoges, France
pascale.magaud@unilim.fr

[3]Laboratoire d'Ingénierie des Systèmes Biologiques et des Procédés (LISBP), Université de Toulouse,
CNRS, INRA, INSA, 135 Avenue de Rangueil, 31077 Toulouse, France
christine.lafforgue@insa-toulouse.fr

## KEY WORDS

suspensions, inertial focusing, particle-laden flows, high Reynolds numbers

## SHORT SUMMARY

The inertial migration of particles in square channel flows at the micro-scale has been deeply investigated in the last two decades. The well-known four equilibrium positions are located near the center of each channel face at moderate Reynolds numbers [1]. More recently, Miura *et al.* [2] revealed experimentally the presence of eight equilibrium positions in millimetric square channels for Reynolds numbers higher than 250. The aim of the present work is to extend these results obtained at millimeter scale to the micrometer scale. To this end, in situ visualization of particles flowing in square micro-channels at Reynolds numbers ranging from 5 to 300 have been conducted and analyzed.

## EXTENDED ABSTRACT

### Introduction

The inertial migration of neutrally buoyant particles was firstly quantitatively highlighted by Segre and Silberberg in the 1960s. Millimetric particles, flowing in a millimeter-scale straight cylindrical tube, were observed to migrate at moderate Reynolds numbers across the streamlines and to focus on an annulus with a radius of ~ 0.6 times the tube radius [3]. However, at this diameter, long channels of several meters were needed for the particles to reach their final equilibrium positions. A few decades later, thanks to the advances in microfabrication techniques, micro-scale channels were used. Since then, the inertial migration phenomenon in microchannel has gained considerable attention due to a wide variety of applications such as separation, concentration, counting, detecting, sorting or mixing of particles in suspension.

Experimental studies with square microchannels have shown a similar migration phenomenon. Di Carlo *et al.* [1] in 2007 observed that neutrally buoyant particles flowing at moderate Reynolds numbers focus only on four equilibrium positions located near each channel wall along their symmetry planes. These results were confirmed by numerical simulations [4, 5]. Square channels are thus particularly advantageous due to their easy manufacturing process and the reduced numbers of focusing positions.

---
\* Corresponding author

However, several experimental and numerical studies have recently concluded that, depending on the flow conditions, the number of equilibrium positions could vary and they could be located in different zones. Abbas *et al.* [5] highlighted that in square micro-channels, particles concentrate in the channel centerline at low flow inertia (typically Reynolds number $Re < 1$). At the same time, Miura *et al.* [2] have obtained experimentally eight equilibrium positions in square millimetric channels for Reynolds numbers higher than 250. These eight equilibrium positions, located at the centers of the channel faces and at the corners were also confirmed numerically by Nakagawa *et al.* [6] in 2015.

Considering the variety of applications of particle inertial migration in microfluidics, it seems important to perfectly identify the different equilibrium positions for a wide range of Reynolds numbers and to clearly evaluate the efficiency of the inertial focusing in these flow conditions. The aim of this work is then to check the validity of Miura *et al.* results at the micrometer scale.

In the present paper, experiments with neutrally buoyant spherical particles flowing in square micro-channels are conducted at Reynolds numbers ranging from 5.6 to 280. A so-called "focusing degree" is introduced to quantify the development of the inertial focusing. Influence of the particle to channel size ratio on the focusing degree are presented. Finally, a possible distribution of the particles in the cross-section for Reynolds numbers higher than 250 is proposed.

**Experimental method**

The spherical particles used for this study are made of polystyrene (Microparticles GmbH®) with diameters $d_p$ = 5.3 and 8.44 µm and a density $\rho$ = 1050 kg m$^{-3}$. The fluid density is matched to the particle one by using a mixture of 23 % glycerol and 77 % deionized water with a particles volume fraction $\Phi$ = 0.1 % for the 8.44 µm particles and $\Phi$ = 0.025 % for the 5.3 µm particles, ensuring the same number of particles in a given volume for both cases. The suspension flows in square borosilicate micro-channels (Vitrocom) with 80 × 80 µm$^2$ inner section and 10 cm length. The flow rates generated by a syringe pump (PDH 4400, Harward Apparatus), range from 667 nL/s to 33 µL/s corresponding to $Re$ from 5.6 to 280 ($Re = UH/\nu$ where $U$ is the mean flow velocity, $H$ the channel height and $\nu$ the kinematic viscosity of the mixture). Particle positions in the channel are in situ observed thanks to an Olympus microscope (BX 51) and images are recorded in several places along the channel length by a high-speed camera (Photron Fastcam SA3) focusing at the midplane of the channel. At least 2000 images are captured to yield satisfactory results. The setup and the image processing have been detailed in previous works [5, 7].

**Results and discussion**

Fig. 1 displays the overall particle distributions for different $Re$. These distributions are obtained by superposing in the same image the centers of particles identified in the 2000 acquired images. For Reynolds number less than 150, the total migration is not complete at a distance from channel inlet $z/H$ = 1000. The particles distributions visualized in the images thus represent the development of the inertial migration. Indeed, in Fig. 1, at $Re$ = 14, particles are still randomly distributed over the image, indicating a relatively uniform distribution in the cross-section. As the Reynolds number increases from 14 to 112, particles are seen to concentrate in lines located near the channel walls and on the centerline of the picture, corresponding to the four well known equilibrium positions. This result confirms that a shorter streamwise length is required for focalization when the Reynolds number is increased [5]. For Reynolds numbers higher than 150, the migration is fully-developed at $z/H$ = 1000. At $Re$ = 280, the middle band which corresponds to the top and the bottom focusing zones is larger as seen in Fig. 1, suggesting that the particles are no longer concentrated in the four equilibrium positions but are distributed more widely at the periphery of the channel cross-section. This decrease in the focalization efficiency might be due to the particle migration towards new additional equilibrium positions near the channel corners, as observed by Miura *et al.* [2] in square millimetric channel for $Re$ higher than 250.

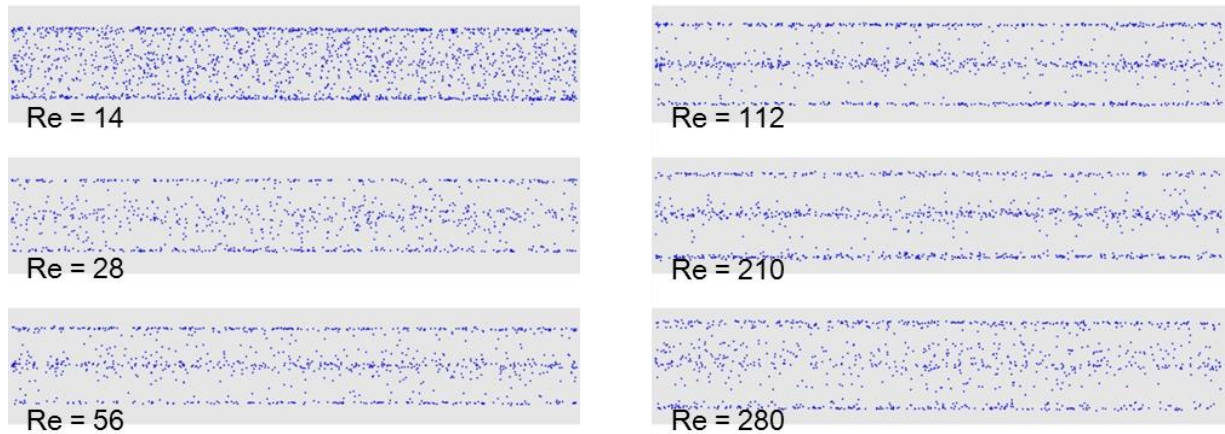

**Figure 1**: Overall particle distributions observed from channel midplane for different *Re* at a distance $z/H = 1000$ from channel inlet and for particles of diameter $d_p = 8.44$ µm.

Since at moderate *Re* in square channels, particles are known to focus at four equilibrium positions near the centers of the channel faces, a parameter $\eta$, termed "focusing degree" and defined as the percentage of particles focused at the four equilibrium positions in comparison with the total number of particles in the microchannel has been introduced [7].

Fig. 2 illustrates the focusing degree $\eta$ for various *Re* at a distance from channel inlet $z/H = 1000$ and for two different particle to channel size ratios. For $d_p/H = 0.11$ (particles of diameter $d_p = 8.44$ µm), $\eta$ increases with increasing *Re*, reaches a maximum at *Re* ~ 150 and then decreases. This confirms that at a fixed distance from channel inlet, the focalization is first enhanced with increasing Reynolds number up to a maximum when the focalization becomes fully-developed. The decrease of $\eta$ observed for higher *Re* could be due to the arising of other equilibrium positions near the channel corners [2].

For $d_p/H = 0.066$, the evolution is similar. However, the maximum percentage of focused particles is lower and is reached at a higher Reynolds number. This result confirms that a higher streamwise length is required for the focalization of smaller particles [5]. In addition, it seems that a higher level of focalization is reached for larger particles.

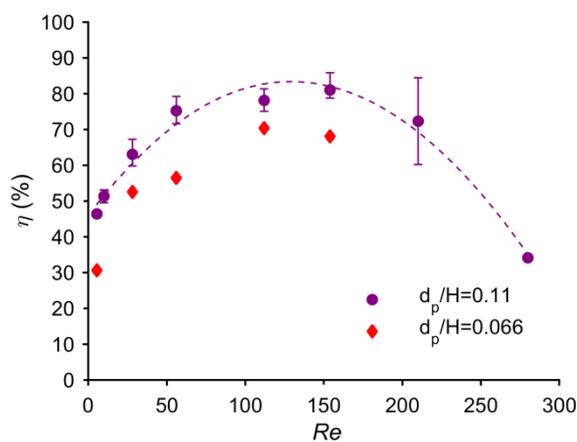

**Figure 2**: Focusing degree $\eta$ at $z/H = 1000$ from channel inlet with respect to *Re,* for different particle to channel size ratios $d_p/H$.

To confirm the existence of the eight equilibrium positions previously observed by Miura, several other experiments have been conducted for Reynolds numbers higher than ~ 250. As shown in Fig. 3 (Left), additional focusing lines appear, suggesting the cross sectional distribution presented in Fig. 3 (Right). Further experiments have however to be done to confirm this trend, for example by using the projection experimental method described in [8] which allows to obtain an image distribution of the particles in the cross section at the outlet of the channel.

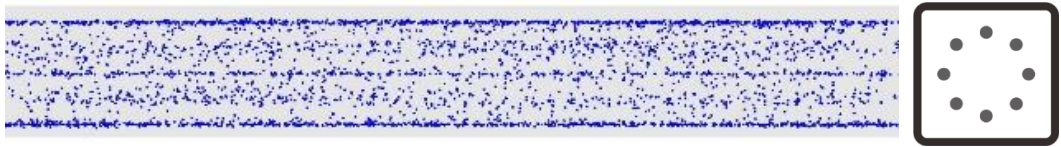

**Figure 3**: **Left**: Overall particle distributions observed from channel midplane for *Re* = 280 at a distance *z/H* = 1000 from channel inlet and for particles with diameter $d_p$ = 8.44 µm in a bidisperse suspension. **Right**: Deduced particle distributions across the channel cross-section.

**Conclusion**

The inertial migration towards the well-known four equilibrium positions of neutrally buoyant particles flowing in square micro-channels could be used to separate, concentrate or detect particles from a mixture. The performance of these devices depends on the efficiency of the inertial migration. We show in this work that at a fixed distance from channel inlet a specific Reynolds number exists for which the focalization is optimal, i.e., for which the rate of particles located on the four equilibrium positions is the highest. Under this value, the migration is not fully-developed and over this value, new additional equilibrium positions arise. This "optimal" Reynolds number depends on the particle to channel size ratio. The determination of this "optimal" Reynolds number is essential for the design of new separation devices.

**Acknowledgements**

The authors acknowledge FERMaT research federation for its support.

**References and Citations**

[1] Di Carlo, D., Irimia, D., Tompkins, R.G., & Toner, M. (2007). Continuous inertial focusing, ordering, and separation of particles in microchannels. *Proceedings of the National Academy of Sciences*, **104**(48), 18892-18897.
[2] Miura, K., Itano, T., & Sugihara-Seki, M. (2014). Inertial migration of neutrally buoyant spheres in a pressure-driven flow through square channels. *Journal of Fluid Mechanics*, **749**, 320-330.
[3] Segre, G., & Silberberg, A. (1961). Radial particle displacements in Poiseuille flow of suspensions. *Nature*, **189**, 209-210.
[4] Di Carlo, D., Edd, J.F., Humphry, K.J., Stone, H.A., & Toner, M. (2009). Particle segregation and dynamics in confined flows. *Phys Rev Lett*, **102**(9), 094503.
[5] Abbas, M., Magaud, P., Gao, Y., & Geoffroy, S. (2014). Migration of finite sized particles in a laminar square channel flow from low to high Reynolds numbers. *Physics of Fluids*, **26**(12), 123301.
[6] Nakagawa, N., Yabu, T., Otomo, R., Kase, A., Makino, M., Itano, T., & Sugihara-Seki, M. (2015). Inertial migration of a spherical particle in laminar square channel flows from low to high Reynolds numbers. *Journal of Fluid Mechanics*, **779**, 776-793.
[7] Gao, Y., Magaud, P., Baldas, L., Lafforgue, C., Abbas, M., & Colin, S. (2017). Self-ordered particle trains in inertial microchannel flows. *Microfluidics and Nanofluidics*, **21**(10), 154.
[8] Lafforgue-Baldas, C., Magaud, P., Schmitz, P., Zhihao, Z., Geoffroy, S., & Abbas, M. (2013). Study of Microfocusing Potentialities to Improve Bioparticle Separation Processes: Towards an Experimental Approach. *Journal of Flow Chemistry*, **3**(3), 92-98.